# Interestingness Measure for Mining Spatial Gene Expression Data using Association Rule

M.Anandhavalli, M.K.Ghose, and K.Gauthaman

**Abstract**—The search for interesting association rules is an important topic in knowledge discovery in spatial gene expression databases. The set of admissible rules for the selected support and confidence thresholds can easily be extracted by algorithms based on support and confidence, such as Apriori. However, they may produce a large number of rules, many of them are uninteresting. The challenge in association rule mining (ARM) essentially becomes one of determining which rules are the most interesting. Association rule interestingness measures are used to help select and rank association rule patterns. Besides support and confidence, there are other interestingness measures, which include generality reliability, peculiarity, novelty, surprisingness, utility, and applicability. In this paper, the application of the interesting measures entropy and variance for association pattern discovery from spatial gene expression data has been studied. In this study the fast mining algorithm has been used which produce candidate itemsets and it spends less time for calculating k-supports of the itemsets with the Boolean matrix pruned, and it scans the database only once and needs less memory space. Experimental results show that using entropy as the measure of interest for the spatial gene expression data has more diverse and interesting rules.

**Index Terms**—Association rule mining, Confidence, Entropy, Interesting Measure, Support, Variance,

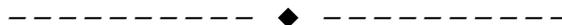

## 1 INTRODUCTION

The process of discovering interesting and unexpected rules from large data sets is known as association rule mining. The typical approach is to make strong simplifying assumptions about the form of the rules, and limit the measure of the rule quality to simple properties such as support or confidence. Support and confidence limit the level of interestingness of the generated rules. Interestingness measures are necessary to help select and rank association rule patterns. Each interestingness measure produces different results, and experts have different opinions of what constitutes a good rule [1]. The interestingness of discovered association rules is an important and active area within data mining research [2]. The primary problem is the selection of interestingness measures for a given application domain. However, there is no formal agreement on a definition for what makes rules interesting. Association rule algorithms produce thousands of rules, many of which are redundant [3, 4]. In order to filter the rules, the user generally supplies a minimum threshold for support and confidence.

Support and confidence are basic measures of association rule interestingness. However, generating rules that meet minimum thresholds for support and confidence may not be interesting. This is because rules are often produced that are already known by a user who is familiar with the application domain. The challenge in association rule mining (ARM) essentially becomes one of determining which rules are the most interesting.

In this paper, an attempt has been made to study the interestingness measure information function entropy and variance for mining association rule from the spatial gene expression data.

## 2 MATERIALS AND METHODS

### 2.1 Spatial Gene Expression Data

The Edinburgh Mouse Atlas gene expression database (EMAGE) is being developed as part of the Mouse Gene Expression Information Resource (MGEIR) [24] in collaboration with the Jackson Laboratory, USA. EMAGE (http://genex.hgu.mrc.ac.uk/Emage/database) is a freely available, curated database of gene expression patterns generated by in situ techniques in the developing mouse embryo. The spatial gene expression data are presented as N×N similarity matrix. Each element in the matrix is a measure of similarity between the corresponding probe pattern and gene-expression region. The similarity is calculated as a fraction of overlap between the two and the total of both areas of the images. This measurement is intuitive, and commonly referred to as the Jaccard index [25]. When a pattern is compared to itself, the Jaccard value is 1 because the two input spatial regions are identical. When it is compared to another pattern, the Jaccard Index will be less than one. If the Jaccard Index is 0, the two patterns do not intersect. If a Jaccard Index value is close to 1, then the two patterns are more similar.

However, biologists are more interested in how gene

---

- *M.Anandhavalli and Dr.M.K.Ghose are with the CSE Department, SMIT, Majitar, East Sikkim.*
- *Dr.K.Gauthama is with the Department of Drug Technology, Higher Institute of Technology, Derna, Libya.*

expression changes under different probe patterns. Thus, these similarity values are discretized such that similarity measure greater than some predetermined thresholds and converted into Boolean matrix.

## 2.2 Data Preprocessing

Preprocessing is often required before applying any data mining algorithms to improve performance of the results. The preprocessing procedures are used to scale the data value either 0 or 1. The values contained in the spatial gene expression matrix had to be transformed into Boolean values by a so-called discretization phase. In our context, each quantitative value has given rise to the effect of discretization procedure [25]: Max minus x% method.

Max minus x% procedure consists of identifying the highest expression value (HV) in the data matrix, and defining a value of 1 for the expression of the gene in the given data when the expression value was above HV – x% of HV where x is an integer value. Otherwise, the expression of the gene was assigned a value of 0 (Table 1).

In the similarity matrix, the items I are genes from the data set, where a transaction T⊂ I consists of genes that all have an expression pattern intersecting with the same probe pattern.

The sets of transactions are constructed by taking, for each probe pattern r, every gene g from which its associated gene expression pattern ge satisfies the minimum similarity β, i.e., similarity(r, ge) > β, to form the itemsets.

TABLE 1
RESULTS OF MAX MINUS 25% DISCRETIZATION METHOD

|   | α (Input) | α (after discretization) |
|---|-----------|--------------------------|
| a | 0.096595  | 0 |
| b | 0.123447  | 0 |
| c | 0.291310  | 1 |
| d | 0.126024  | 0 |
| e | 0.155819  | 0 |
| f | 0.288394  | 1 |
| g | 0.000000  | 0 |
| h | 0.215049  | 1 |

## 2.3 Association Rule Mining

Association rules were first proposed by Agrawal et al. [5]. Agrawal et al.'s work established a formal model for association rules and establishes algorithms that find large itemsets, confidence, and support of each rule discovered in the itemset. A primary goal of knowledge discovery in databases is to produce interesting rules that can be interpreted by a user [6]. Association rules are used widely in the retail industry under the name 'market basket analysis'. Association rules have been used as well to mine medical record data [7], [8]. The general definition for association rules is as follows:

An 'association rule' is a pair of disjoint itemsets. If LHS and RHS denote the two disjoint itemsets, the association rule is written as LHS→RHS i.e LHS and RHS are sets of items, the RHS set being likely to occur whenever the LHS set occurs.

The 'support' of the association rule LHS→RHS with respect to a transaction set T is

$$SUP(A \cup B) = P_{ab} = \frac{n_{ab}}{n}$$

The 'confidence' of the rule LHS→RHS with respect to a transaction set T is

$$CONF(A \cup B) = \frac{P_{ab}}{P_a} = \frac{n_{ab}}{n} = 1 - \frac{n_{a\bar{b}}}{n_a}$$

In market basket analysis, an association rule represents a set of items that are likely to be purchased together; for example, the rule {cereal}→{milk, juice} would state that whenever a customer purchases cereal, he or she is likely to purchase both milk and juice as well in the same transaction. In the analysis of gene expression data, the items in an association rule can represent genes that are strongly expressed or repressed, as well as relevant facts describing the cellular environment of the genes (e.g. a diagnosis for a tumour sample that was profiled, or a drug treatment given to cells in the sample before profiling). An example of an association rule mined from expression data might be {cancer} → {gene A↑, gene B↓, gene C↑}, meaning that, for the data set that was mined, in most profile experiments where the cells used were cancerous, gene A was measured as being up (i.e. highly expressed), gene B was down (i.e. highly repressed), and gene C was up, altogether.

## 2.4 Interestingness Measures

Interestingness measures can be classified into two categories: objective and subjective. Objective measures are based on statistics, while subjective measures are based on an understanding of the user's domain knowledge. For example, objective measures include generality and reliability, conciseness, peculiarity, diversity and surprisingness. The subjective interestingness measures include novelty, utility, and applicability. These measures assist in validating association rule results [22].

What makes a rule interesting? One way to define interestingness is that a rule must be valid, new and comprehensive [23]. Commonly used measures of interestingness include support and confidence as described in Table 2.

Two important measures within association rule mining are support and confidence. Support for an association rule is the percentage of transactions in the database that contain X U Y. Confidence for an association rule (sometimes denoted as strength, or α) is the ratio of the number of transactions that contain XUY to the number of transactions that contain X [9]. A third measure of interestingness is lift. Lift is a measure of the probability of finding the consequent in any random basket. In other words, lift "measures how well the associative rule performs by comparing its performance to the "null" rule" [10]. In this study, interesting measure Shannon's entropy has been considered, because of their wide-spread use. It is also important to note that there are several different measures for entropy, which is a mathematical measure of information loss. Generally, we try to minimize this loss, so lower values would indicate less in-formation loss.

TABLE 2
LIST OF MEASURES OF INTEREST

| Measure | Measure | Acronym | Ref. |
|---|---|---|---|
| Support | $P_{ab}$ | SUP | [5] |
| Confidence | $P_{a/b}$ | CONF | [5] |
| Lift | $\dfrac{P_{ab}}{P_a P_b}$ | LIFT | [11] |
| Ganascia | $2P_{ab} - 1$ | GAN | [12] |
| Piatetsky-Shapiro | $nP_a (P_{a/b} - P_{a/\bar{b}})$ | PS | [13] |
| Loevinger | $\dfrac{P_{b/a} - P_{b/\bar{a}}}{P_{\bar{b}}}$ | LOE | [14] |
| Zhang | $\dfrac{P_{ab} - P_a P_b}{\max\{P_{ab} P_{\bar{b}}; P_b P_{a\bar{b}}\}}$ | ZHANG | [15] |
| Implication Index | $\sqrt{n}\, \dfrac{P_{a\bar{b}} P_a P_{\bar{b}}}{\sqrt{P_a P_{\bar{b}}}}$ | IMPIND | [16] |
| Least contradiction | $\dfrac{P_{ab} P_{a\bar{b}}}{P_b}$ | LC | [17] |
| Conviction | $\dfrac{P_a P_{\bar{b}}}{P_{a\bar{b}}}$ | CONV | [18] |
| Implication Intensity | $P[\text{Poisson}(nP_a P_{\bar{b}}) \geq nP_{a\bar{b}}]$ | IMPINT | [10] |
| Sebag-Schoenauer | $\dfrac{P_{ab}}{P_{a\bar{b}}}$ | SEB | [20] |
| Bayes Factor | $\dfrac{P_{ab} P_{\bar{b}}}{P_{a\bar{b}} P_b}$ | BF | [21] |

### 2.4.1 Entropy and Variance

Shannon's entropy measure is based on information theoretic function and entropy. It measures the relative information content present in a dataset. The Shannon's measure is calculated using the formula:

$$\text{Entropy} = -\sum_{i=1}^{n} P_i \log_2 P_i \quad \ldots (1)$$

Variance is a simple measure that can be used to compare two data sets and the rule diversity of each data set. The variance measure is calculated using the formula:

$$\text{variance} = \dfrac{\sum_{i=1}^{n}(P_i - \bar{q})^2}{n-1} \quad \ldots (2)$$

In equation (1) and (2), Pi is the probability for the class i and $\bar{q}$ is the average probability for all class.
A higher variance indicates that the rules are more diverse. The major challenge in deciding which data set is more diverse is that different measures produce different values.

### 2.5 Method details

The method for finding the interesting measure of spatial gene expression data in the form of similarity matrix consists of the following steps:
1. Apply fast association rule mining algorithm for generating the frequent itemsets and association rules from the spatial gene expression dataset.
2. List all association rules along with the measures support, confidence.
3. Filter the top 15 association rules ranked in order by confidence value.
4. Apply entropy and variance measure to determine relative interestingness.

A detailed description of the introduced algorithm for generating frequent itemsets and association rules is as follows:

*Input: Spatial Gene Expression data in similarity matrix (M), the minimum support.*
*Output: Set of frequent itemsets F.*

1. Normalize the data matrix M and transformed into Boolean Matrix B;
   // Frequent 1-itemset generation
2. for each column $C_i$ of B
3. If $sum(C_i)$ >= new_support
4. F1 = { $I_i$};
5. Else delete $C_i$ from B;
6. for each row $R_j$ of B
7. If $sum(R_j)$ < 2
8. Delete $R_j$ from B;
9. for (k=2; | $F_{k-1}$ | > k-1; k++)
10. {
11. Produce k-vectors combination for all columns of B;
12. for each k-vectors combination { $B_{i1}, B_{i2},…B_{ik}$}
13. { E= $B_{i1} \cap B_{i2} \cap …\cap B_{ik}$
14. If sum(E) >= new_support
15. Fk = { $I_{i1}, I_{i2},…I_{ik}$}
16. }
17. for each item Ii in Fk
18. If | $F_k(Ii)$ | < k
19. Delete the column Bi according to item Ii from B;
20. for each row Rj of B
21. If $sum(B_j)$ < k+1
22. Delete $B_j$ from B;
23. k=k+1
24. }
25. Return F = $F_1$ U $F_2$….U $F_k$.
    // List all association rules along with the measures support, confidence
26. for all $f_k$, $f_k$ ∈ F, 1<=k<=maxsize-1 do begin
27. rsup=support($f_k$)*miconf
28. found=0
29. for all fm, fm ∈ F k+1<= m <=maxsize do begin
30. if (support($f_m$)>=rsup) then begin
31. if($f_k \subset f_m$) then begin
32. found=found+1
33. conf=support($f_m$)/ support($f_k$)
34. generate the rule $f_k$ = ( $f_m$- $f_k$) &= conf and support=support($f_m$)

35. *end if*
36. *else*
37. *if (found<2)*
38. *continue step1 with next k*
39. *else found=0*
40. *endif*
41. *endif*
42. *end do*
43. *end do*
44. *List the top 10 association rules ranked in order by confidence value.*
45. *Apply entropy and variance measure to determine relative interestingness.*

## 3 PRELIMINARY RESULTS AND DISCUSSION

The primary goal of the spatial gene expression data set is to find the set of genes which expressed very highly. In this study, we are using association rules to help make predictions.

TABLE 3
RESULTS OF ASSOCIATION MINING FROM STAGE 14 OF EMAGE DATABASE

| ************ | | |
|---|---|---|
| 15 most confident rules | | |
| Confidence | Support | Rules |
| 1 | 0.05 | Hist1h2bc (EMAGE:1218) -> Bmp7( EMAGE:621) |
| 1 | 0.06 | {T (EMAGE:1064), Cdx2 (EMAGE:3327)} -> Hoxc8 (EMAGE:905) |
| 1 | 0.07 | Hist1h2bc (EMAGE:1218) -> Bmp7( EMAGE:621) |
| 1 | 0.05 | Hoxc8 (EMAGE:902) -> Hoxc8 (EMAGE:905) |
| 1 | 0.06 | T (EMAGE:1064)-> Hoxc8 (EMAGE:902) |
| 1 | 0.07 | Vezf1EMAGE:3229 -> Pecam1EMAGE:3384 |
| 1 | 0.05 | Snai2(EMAGE:3236)-> Snai2 (EMAGE:3239) |
| 1 | 0.06 | T (EMAGE:1064)-> Hoxc8 (EMAGE:905) |
| 1 | 0.07 | Vezf1EMAGE:3229 -> Bmp7(EMAGE:621) |
| 1 | 0.05 | {T (EMAGE:1064), Cdx2 (EMAGE:3327)} -> Hoxc8 (EMAGE:905) |
| 1 | 0.07 | Pecam1EMAGE:3384->Bmp7( EMAGE:621) |
| 1 | 0.08 | Hoxc8 (EMAGE:902) -> Hoxc8 (EMAGE:905) |
| 1 | 0.08 | Hist1h2bc (EMAGE:1218) -> Bmp7( EMAGE:621) |
| 1 | 0.09 | Hoxc8 (EMAGE:902) -> Hoxc8 (EMAGE:905) |
| 1 | 0.09 | Hist1h2bc (EMAGE:1218) -> Bmp7( EMAGE:621) |
| ************ | | |

The dataset showed 15 rules ranked in order by confidence. Support ranges from 0.05 to 0.09. The value of 5% means that the algorithm found 5% of the transactions contained the left hand side of the rule. For example, in examining the first rule produced from the data set, we find 5% of expressed patterns contains gene Hist1h2bc (EMAGE:1218) in Stage 14 of EMAGE database given in Table 3.

From these preliminary results, spatial gene expression data would be considered more interesting based on the entropy. A lower variance (0.0001) indicates that the rules are not interesting. The major challenge in deciding which data set is more diverse is that different measures produce different values. Analysis of the spatial gene expression data provided an entropy measure of 0.2586. This is expected since lower entropy measures show less information loss. Using entropy as the measure of interest, the spatial gene expression data has more diverse rules. This is encouraging since these measures tell us different things about the rules.

## 4 CONCLUSION

In this paper, a novel method of finding interesting measure for the spatial gene expression data has been studied. The fast association rule mining algorithm used in this study, which does not produce candidate itemsets and it spends less time for calculating k-supports of the itemsets with the Boolean matrix pruned, and it scans the database only once and needs less memory space when compared with Apriori algorithm. From the study, it has been found that using entropy as the measure of interest for the spatial gene expression data has more diverse and interesting rules. Finally, the large and rapidly increasing compendium of data demands data mining approaches, particularly association rule mining with various interesting measure ensures that genomic data mining will continue to be a necessary and highly productive field for the foreseeable future.

## ACKNOWLEDGMENT

This study has been carried out as part of Research Promotion Scheme (RPS) Project under AICTE, Govt. of India.